\documentclass[runningheads]{llncs}

\usepackage{graphicx}

\usepackage{xcolor}
\usepackage[hidelinks]{hyperref}

\usepackage{amsmath}
\usepackage{amssymb}
\usepackage{cleveref}
\usepackage{caption}
\usepackage{subcaption}
\usepackage{multirow}

\usepackage[ruled,linesnumbered]{algorithm2e}
\DontPrintSemicolon

\crefname{algocf}{algorithm}{algorithms}
\Crefname{algocf}{Algorithm}{Algorithms}

\makeatletter
\DeclareRobustCommand\ttfamily
        {\not@math@alphabet\ttfamily\mathtt
         \fontfamily\ttdefault\small\selectfont}
\makeatother

\input{listings}

\begin{document}

\title{Combining Theory of Mind and Abduction for Cooperation under Imperfect Information}

\titlerunning{Combining Theory of Mind and Abductive Reasoning}

\author{Nieves Montes \and
Nardine Osman \and
Carles Sierra}

\authorrunning{N. Montes et al.}

\institute{
 Artificial Intelligence Research Institute (IIIA-CSIC) \\
 08193 Bellaterra, Barcelona (Spain) \\
 \email{\{nmontes,nardine,sierra\}@iiia.csic.es}
}

\maketitle

\begin{abstract}
In this paper, we formalise and implement an agent model for cooperation under imperfect information. It is based on Theory of Mind (the cognitive ability to understand the mental state of others) and abductive reasoning (the inference paradigm that computes explanations from observations). The combination of these two techniques allows agents to derive the motives behind the actions of their peers, and incorporate this knowledge into their own decision-making. We have implemented this model in a totally domain-independent fashion and successfully tested it for the cooperative card game Hanabi.

\end{abstract}

\section{Introduction}\label{sec:intro}
The emergent field of social AI deals with the theoretical foundations and practical implementations of autonomous agents that are able to interact with other agents, possibly including humans \cite{Dafoe2021}. In order for autonomous agents to be socially competent, they must take into account not only their own goals and point of view, but also those of their fellow agents. The cognitive ability to put oneself in the shoes of someone else and reason from their perspective is called {\em Theory of Mind} (ToM). In order to have software agents with ToM capabilities, they must explicitly incorporate some technique for {\em modelling others}, the preferred term within the AI community \cite{Albrecht2018}.

Techniques for modelling others are fairly prevalent within AI, particularly in competitive domains characterised by agents with diverging interests \cite{Baarslag2015,Nashed2022}. However, endowing agents with ToM faculties has the potential to boost their performance in cooperative tasks too, where agents must collaborate with one another in an efficient manner. In particular, in domains dealing with imperfect information (i.e. where agents do not have access to the complete state of the system but can infer the subset of states that are currently possible), autonomous agents can benefit from observing the actions performed by others, inferring the knowledge their peers were relying upon when selecting their actions, and incorporating this additional information into their own decision-making. This type of reverse inference, from observations to potential premises, is called {\em abduction} and is central to the work presented here.

In this paper, we present preliminary work on the formulation and implementation of an agent model combining Theory of Mind and abductive reasoning capabilities in purely cooperative tasks characterised by imperfect information. To cope with this, we propose a framework for agents that observe the actions of their teammates, adopt their perspective (thus utilising Theory of Mind) and generate explanations concerning the knowledge they were relying upon to decide on that action (hence engaging in abductive reasoning). We also review how agents update and incorporate these explanations into their knowledge base for their own decision-making.

Although we provide a tentative decision-making procedure for action selection that takes into account abductive explanations, this is not the main contribution of this work. The focus of this work is the derivation of knowledge using ToM and abductive reasoning. The ways in which agents use such knowledge for strategy selection (the primary concern of the epistemic game theory literature \cite{Perea2012}) is a necessary component of the overall agent software, but a thorough examination of it is beyond the scope of this work. Furthermore, it should be noted that our model is completely domain-independent, but works under some broad assumptions that we specify.

Our work can be compared with previous approaches in the plan recognition literature \cite{VanHorenbeke2021}, where agents infer the goal and sequence of actions (i.e. the \emph{plan}) that others are pursuing, in order to anticipate and respond to future actions. In this paper, we do not work with plans \emph{per se} that include sequences of actions, but with atomic actions. Consequently, we are not interested in identifying a full sequence of actions under execution, but the circumstances that have led to an action choice. This is analogous to the recognition of the plan \emph{context} in many BDI languages \cite{Mascardi2005}.

This paper is organised as follows. First, \Cref{sec:background} provides the necessary background on Theory of Mind and abductive reasoning. \Cref{sec:background} also presents the cooperative game Hanabi, which we will be using as our running example throughout the exposition of our agent model in \Cref{sec:combining}. Although the model we provide is totally domain-independent, illustrating it with a running example provides a much clearer picture. Finally, \Cref{sec:results} presents results on the performance of our agent model, and we conclude in \Cref{sec:conclusions}.

\section{Background}\label{sec:background}
We start by providing an overview of the two techniques that our agent model combines: Theory of Mind (ToM) and abductive reasoning. Both terms are relevant in many fields beyond AI, and hence the precise use we make of them here needs clarification. These introductions are fairly general. The specific way in which we use ToM and abductive reasoning is covered in \Cref{sec:combining}.

\subsection{Theory of Mind}\label{subsec:tom}
Theory of Mind (ToM) refers to the human cognitive ability to perceive and understand others in terms of their mental attitudes, such as their beliefs, emotions, desires and intentions \cite{Malle2022}. Humans routinely interpret the behaviour of others in terms of their mental states, and this capacity is considered essential for successful participation in social life.

From the philosophical and psychological perspectives, there are two disparate views on Theory of Mind: Theory ToM (TT) and Simulation ToM (ST) \cite{RoeskaHardy}. Theory ToM argues that the attribution of mental states to others happens according to internally represented knowledge, analogous to a theory of folk psychology. This theory is implicit, and is so pervasive and integral to our lives that it goes unnoticed. In contrast, Simulation ToM takes the view that one uses one's own mind as a model to understand the mind of others, with no theoretical knowledge involved. Instead, one puts oneself in the shoes of others by pretending to be in their circumstances, and performs a sort of mental simulation using one's own mental mechanisms to predict the thoughts and actions of others. In this work, we adhere closer to the ST account than to the TT one, since ST presents a much clearer path to being operational.

Within AI, ToM has been applied in a somewhat fragmented way, with many fields implementing it based on their prevalent techniques. In machine learning, for example, ToM has been conceived as a meta-learning process \cite{Rabinowitz2018}, where a Deep Neural Network model takes as input past agent trajectories and outputs behaviour at the next time-step. ToM approaches have also been investigated from the perspective of game theory. In \cite{Weerd2011,Weerd2012}, the authors consistently prove that the marginal benefits of employing higher-order ToM (I know that my opponent knows that I know that my opponent knows...) diminish with the recursion level employed. In particular, while first (I know that my opponent knows) and second-order ToM (I know that my opponent knows that I know) present a clear advantage, deeper recursion levels do not.

Finally, purely symbolic approaches to ToM have studied the effect of announcements on the beliefs of others and their ripple-down effects on their subsequent desires and actions. These approaches use modal operators such as $K_i$, $B_i$, $D_i$ to designate the knowledge, beliefs and desires, respectively, of agent $i$. ToM comes into play when such operators are nested within one another, e.g. $K_i K_j \phi$ indicates that $i$ knows that $j$ knows that $\phi$ holds true, a first-order ToM statement. In \cite{Panisson2018,Sarkadi2019}, authors formalise and implement ToM capabilities into symbolic agents for the purposes of deception and manipulation.

\subsection{Abductive Reasoning}\label{subsec:abduction}
In the symbolic AI literature, centre stage has traditionally been taken by deductive reasoning, based on the application of the {\em modus ponens} rule: from knowledge of $\phi$ and $\phi \rightarrow \psi$, infer $\psi$. In contrast to deduction, {\em abductive} reasoning works in the opposite direction: upon knowledge of $\phi \rightarrow \psi$ and the observation of $\psi$, $\phi$ is inferred as a potential {\em explanation} for $\psi$ \cite{Douven2021}.

At a very high level, abduction takes as input (1) a logical theory representing expert knowledge on the domain of interest; and (2) a query in the form of a logical formula that stands for an observation made in that domain. Then, abductive inference computes an explanation formula that, together with the original logical theory, entails the observation and is logically consistent with it.

Computationally, an Abductive Logic Programming (ALP) theory \cite{Denecker2002} is a tuple $\langle T, A \rangle$, where $T$ is a logic program and $A$ is a set of ground abducible atoms.\footnote{Under the restriction that no predicate in $A$ appears as the head of a clause in $T$.} Then, an abductive explanation is defined as follows:

\begin{definition}\label{def:abductive-explanation}
(from \cite{Kakas1992}) Given an ALP theory $\langle T, A \rangle$ and a query $Q$, an {\em abductive explanation} $\Delta$ is a subset of abducible atoms such that $T \cup \Delta \models Q$.
\end{definition}

Often, an ALP theory is extended with a set of integrity constraints (ICs). Then, \Cref{def:abductive-explanation} has to be extended to account for the consistency of $\Delta$ with the ICs. One account for this consistency imposes that $T \cup \Delta \cup IC$ must not lead to contradiction. In our agent model, although we do not use ICs in the traditional sense, we adopt a notion analogous to this consistency view. This allows us to work with incomplete explanations, i.e. $\Delta$ does not necessarily complement the current knowledge to provide a complete representation of the current state, but nonetheless provides valuable information.

In practice, explanations in ALP are computed by extending Selective Linear Definite (SLD) resolution or its negation as failure version (SLDNF) \cite{Denecker1998}. The basic idea is that before failing a goal if one subgoal does not unify with a clause, it should be considered as part of a potential explanation, as long as it unifies with an element in $A$. Hence, goals only fail if they are not provable either by the knowledge base or by matching with the set of abducible atoms. Just as standard SLD(NF) are coupled with backtracking to find all the unifications to a query, so are their abductive counterparts backtracking to find all the possible explanations that render an observation true.

In order for an abductive explanation $\Delta$ to be useful, it needs to be assimilated into the agent's knowledge base (KB). Note that this KB does not necessarily correspond to the logic program $T$ in the ALP theory used for computing the abductive explanations in the first place. Several possibilities may arise when integrating $\Delta$ into a KB \cite{Kakas1992}: (1) the explanation may be uninformative ($KB \models \Delta$); (2) the explanation may render a portion of the knowledge base irrelevant ($KB=KB_1 \cup KB_2$, where $KB_1 \cup \Delta \models KB_2$); (3) $\Delta$ violates the consistency of $KB$, $KB \cup \Delta \models \bot$); and (4) $\Delta$ is independent of $KB$. Of these four possibilities, (4) is clearly the most readily actionable, as it provides additional knowledge to the agent without compromising previously acquired information.

\subsection{The Hanabi Game}\label{subsec:hanabi-game}
We use the award-winning Hanabi game as the test-bed of our agent model, and we will also use it to exemplify the various components in \Cref{sec:combining}. Hanabi is a card game where a team of 2 to 5 players work together with the goal of achieving the maximum possible team score. Every player is handed four or five cards (depending on the size of the team) such that everyone else can see their cards except the player holding them. Every card has a rank between 1 and 5 and one of five colours.\footnote{A detailed description of the rules of the game can be found at \url{https://github.com/Hanabi-Live/hanabi-live/blob/main/docs/RULES.md}.}

\begin{figure}[ht]
    \centering
    \includegraphics[width=0.7\textwidth]{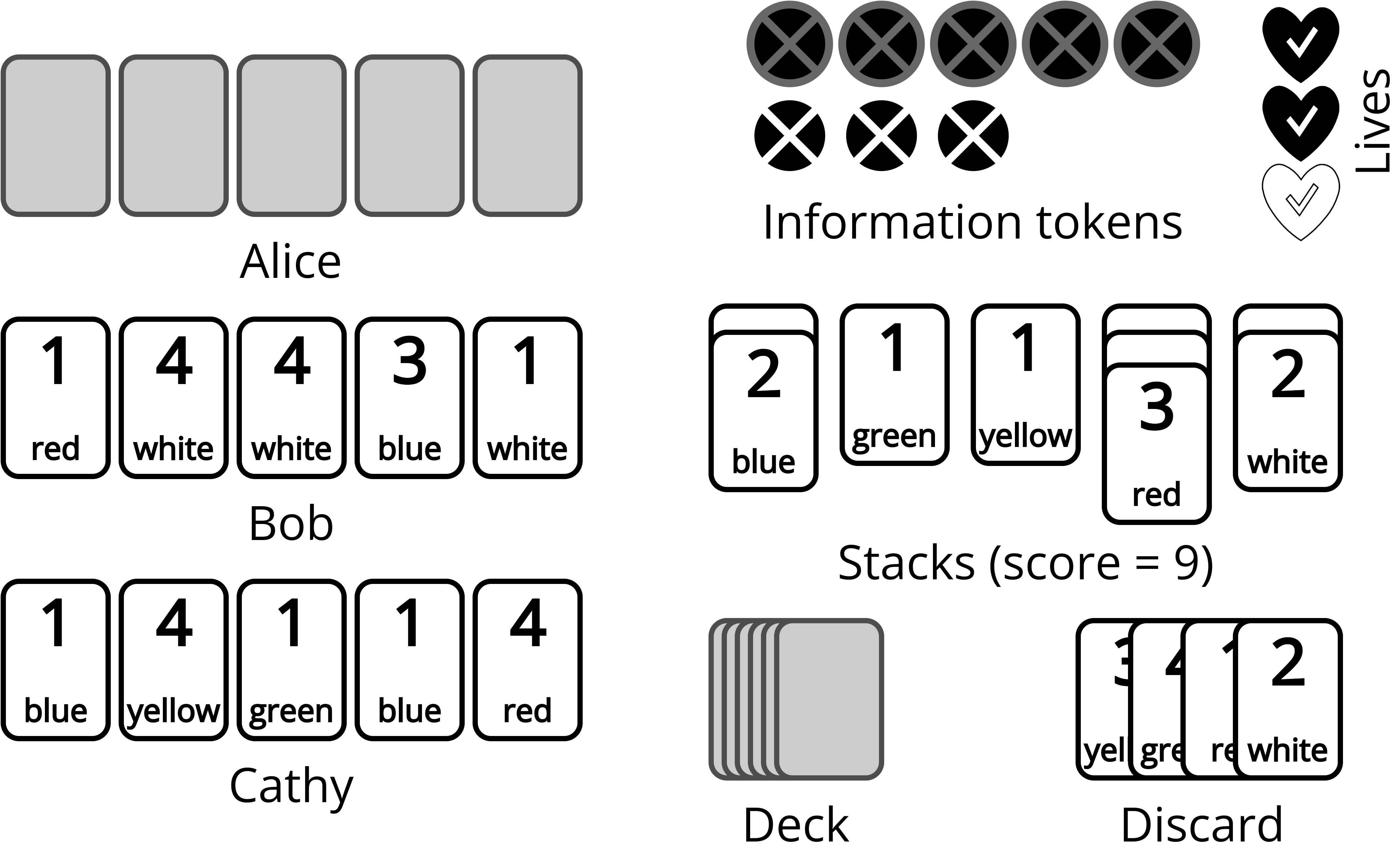}
    \caption{Hanabi game setting for three players, from the perspective of Alice.}
    \label{fig:hanabi-setting}
\end{figure}

The set-up of a typical Hanabi game from the perspective of player Alice is presented in \Cref{fig:hanabi-setting}. To make progress, players take turns to build ordered stacks of cards of the same colour. For example, the red stack is built by playing first a red 1, then a red 2 on top of it, and so on until the red 5. At every turn, players can: (1) play a card on the stacks; (2) discard a card; or (3) give a hint to another player by spending one information token (there are 8 tokens initially available). When playing a card, a participant places it on the stack of the corresponding colour. If the card is not correctly played (e.g. a blue 4 is played when the top of the blue stack has a 2), the whole team lose one life. When a player discards a card, they get rid of it and recover one information token for the team. After playing or discarding a card, players draw a new one from the deck. Finally, players can spend one information token, if available, and give a hint to another player. Players can hint to one another about the rank or the colour of their cards. For example, if Alice hints ``white'' to Bob in \Cref{fig:hanabi-setting}, she must point to the cards on Bob's second, third and fifth slot (starting from the left). Hints are necessarily truthful, as Hanabi is a collaborative game and everyone would lose points by conveying or believing false information. If all information tokens are spent, the player with the turn to move cannot give another hint. The game finishes when the team lose all three lives and get a score of 0, when they manage to finish all of the stacks and get the maximum score of 25, or when they run out of cards to draw from the deck. The final score corresponds to the sum of the top cards in each stack.

There are three main features of Hanabi that make it an excellent test-bed to assess techniques for modelling others in cooperative domains, and that have led some researchers to point to Hanabi as the next grand challenge to be tackled by the AI community \cite{Bard2020}. First, Hanabi is a purely cooperative game. This means that participants can greatly benefit from understanding the mental states of others, e.g. their goals and intentions, in order to align their own actions with those of their teammates.

Second, players in Hanabi must deal with {\em imperfect} information, as they can see the cards of others but not their own. To cope with this, agents must provide information to one another through hints. There are two facets to this information: the explicit knowledge carried by the hint (i.e. the colour or rank of the cards involved) and the implicit information derived from understanding the player's reasons to make that hint. For example, in \Cref{fig:hanabi-setting}, Alice might hint ``red'' to Cathy, hoping that Cathy will understand that Alice would only provide such a hint if she wanted Cathy to play that card, concluding that, since the card is red, it must be a 4. In our agent model, this implicit information is identified with the abductive explanations that agents compute when they take the perspective of the acting agent and derive additional knowledge from it.

The final interesting feature of Hanabi comes from the fact that information sharing is handled as a collective limited resource. The number of information tokens available must be managed by the whole team, by spending or recovering them, and the total number of tokens is finite.

Previous research on Hanabi-playing agents has, for the most part, adhered to one of two approaches: reinforcement learning (RL) and rule-based agents. In the first case, Hanabi-playing bots are trained using state-of-the-art learning algorithms \cite{Bard2020}. In the second case, agents play according to a set of pre-coded rules that indicate what action to take as a function of the game history and the current state \cite{Cox2015,Eger2017,Bergh2017}. This is the path that we adhere to in this work, since our agent model relies on the assumption of a pre-coded strategic convention being followed by all teammates. However, we will not go into the details of the particular action selection clauses, since our agent model is agnostic with respect to the specifics of the team strategy.

Interestingly, in a recent survey where software agents were paired with human teammates, RL agents were perceived as more unreliable, difficult to understand and overall worse teammates that rule-based agents \cite{Siu2021}. The current state of the art for Hanabi AI combines both RL and pre-coded rules \cite{Lerer2020}. There, the authors first perform single-agent learning, where they fix the strategy to be followed by all players except one learning agent. All other members of the team act according to the same pre-coded rules. Second, they implement multiagent learning, where all agents perform the same policy update after every action, if feasible (if not they fall back on a set of pre-coded rules), so the learned policy is always maintained as common knowledge.

\section{Agent Model}\label{sec:combining}
In this section, we present the agent model combining Theory of Mind and abductive reasoning for cooperation under imperfect information. This model applies to all members in a team modelled as a multiagent system (MAS), which we define as follows:
\begin{definition}\label{def:mas}
A multiagent system is a tuple $\langle G, \mathcal{S}, \mathcal{L}, \mathcal{A} \rangle$ where $G$ is a set of agents; $\mathcal{S}$ is a set of global states; $\mathcal{L}$ is a (first-order) logical language used to describe the domain; $\mathcal{A}$ is a set of agent actions.
\end{definition}

We denote the current state of the system by $\mathbf{s}$, formally specified by a set of ground literals composed of the symbols in $\mathcal{L}$. In general, agents do not have access to all of $\mathbf{s}$, but to a partial representation of it. We denote agent $i$'s current view of $\mathbf{s}$ by $s^i$, also specified by a set of ground literals.

Although \Cref{def:mas} is very general, this work is restricted to a particular subclass of problems, which we refer to as {\em common expertise domains}:
\begin{definition}\label{def:common-expertise}
A {\em common expertise domain} corresponds to a MAS $\langle G, \mathcal{S}, \mathcal{L}, \mathcal{A} \rangle$ where the following properties hold:
\begin{enumerate}
    \item \textbf{Common ontology assumption}: All agents share the same complete ontology about the domain at hand. Agents know about all the features and the possible values that characterise a state. As a consequence, given an arbitrary representation of a state $s^i$, any agent can deduce the subset of states $\mathcal{S}^i \subseteq \mathcal{S}$ that are compatible with $s^i$.
    \item \textbf{Common group strategy}: The group strategy is defined as a mapping $\mathsf{Str}: \mathcal{S} \times G \rightarrow \mathcal{A}^{\mid G \mid}$ of states to the action that should be performed by every agent. We assume that the group strategy function is known to all members of the team.
    \item \textbf{Non-faulty perception}: The information an agent perceives about the current state $\mathbf{s}$ is true, i.e. $s^i \subseteq \mathbf{s}$. Additionally, at every state $\mathbf{s}$, agents reliably perceive actions performed by all other agents, i.e. the tuples $\langle j, a_j \rangle$, $\forall j \in G$.
\end{enumerate}
\end{definition}

The first feature of common expertise domains reflects the idea that all agents approach the task at hand according to a consensus mainstream theory. For instance, the evolution of species and the theory of continental drift are consensus theories in biology and geology, uncontested in the academic world. Additionally, because the ontology that all team members share is complete, they know the features that states are characterised by, even if they cannot observe their values. Hence, agents have the benefit of {\em complete} information, although they still have to deal with {\em imperfect} information. Consequently, agents can infer which states they could be in given their current view of the system.

The second feature entails that all agents can safely assume that everyone else is behaving according to the same set of rules. Necessarily, the team of agents, prior to embarking in the current task, have all agreed on what team strategy to follow. Since we are dealing with cooperative tasks, it is reasonable to assume that agents do not expect any gains from free-riding and deviating from the team strategy, since they benefit from coordinating with one another. Agents extract knowledge from the actions performed by teammates by assuming that they are all following the same strategy. An agent could harm the team if they were to follow a different strategy, thus leading teammates to an erroneous interpretation of their actions. In this paper, we do not investigate how such a strategic convention has come to be selected. We simply assume that such an agreement exists and that it is willingly adopted by all participants. This assumption implies that we are not concerned with synthesising the optimal team strategy. Learning an optimal strategy given the reasoning scheme we present here is a task to be implemented on top of the current agent model, and it is outside the scope of this work.

The third point in \Cref{def:common-expertise} makes our approach squarely a {\em knowledge-based} one, as opposed to a belief-base one, since beliefs are not guaranteed to be true. Beyond the perception of the current state, agents also correctly sense the actions being performed by their teammates. They will need to rely on this knowledge for the abduction task, presented in \Cref{subsec:abduction-task}.

The symbols in $\mathcal{L}$ are used to construct the logical program that every agent operates by. For agent $i$, we denote their program as $T_i$. $T_i$ is composed of the following components:
\begin{itemize}
    \item A set of atoms corresponding to $i$'s current view of $\mathbf{s}$, i.e. $s^i$. For example, in the Hanabi game agent Alice can see Bob's cards, and her view would include literals such as \texttt{has\_card\_color(bob,4,blue)} and \texttt{has\_card\_rank(bob,4,3)}, to indicate that Bob has a blue 3 in his $4^{\text{th}}$ slot.
    \item A set of clauses of the form $h\;\texttt{:-}\;b_1, ..., b_n$. We classify the clauses in $T_i$ into the following categories:
    \begin{itemize}
        \item {\em Domain-related clauses} provide definitions about the domain. For example:
\begin{lstlisting}[language=Prolog]
playable(C,R) :- colour(C), rank(R), stack(C,S), S=R-1.
\end{lstlisting}
        expresses that a card can be correctly played if the stack of the corresponding colour is exactly one level below the rank of the card.
        \item {\em Impossibility constraints} are clauses with atom \texttt{imp} as their head and whose body contains literals that cannot hold simultaneously true. Together, the domain-related clauses and these constraints encapsulate the ontology of the domain at hand, and are shared among all agents in the team. For example:
\begin{lstlisting}[language=Prolog]
imp :- has_card_colour(P,S,C1), has_card_colour(P,S,C2), C1\==C2.
\end{lstlisting}
        states that a player cannot simultaneously hold cards of different colours in the same slot \texttt{S}.
        \item {\em Theory of Mind clauses}, with head \texttt{knows(Agent, Fact)}, express that agent $i$ knows that \texttt{Agent} knows that \texttt{Fact} holds true. We impose the restriction that the \texttt{Fact} argument in the head literal must also appear in the clause body, i.e. \texttt{knows(Agent, Fact) :- ..., Fact, ...} . ToM clauses are called upon when agent $i$ switches to the perspective of agent $j$. However, they play no part when, once $i$ has adopted the perspective of $j$, $i$ \emph{reasons} from $j$'s perspective. That is handled by the abductive reasoning process covered in \Cref{subsec:abduction-task}. This restriction guarantees that $i$ cannot know that $j$ knows something if $i$ does not know about it in the first place nor has $i$ bothered to reason from $j$'s perspective. In epistemic logic notation, this is expressed by the axiom:
        \begin{equation}\label{eq:MAS-axioms}
        	\sim K_i \phi \rightarrow \sim K_i K_j \phi
        \end{equation}
        For example:
\begin{lstlisting}[language=Prolog]
knows(Agj, has_card_color(Agk,S,C)) :-
    has_card_color(Agk,S,C), Agj\==Agk.
\end{lstlisting}
        indicates (from $i$'s perspective) that agent \texttt{Agj} can see the colour of the cards that any other agent \texttt{Agk} has.
        \item {\em Abducible clauses}, with head \texttt{abducible(Fact)}, express all the information that could potentially be added to $i$'s current perception of the environment $s^i$ to reconstruct the complete state $\mathbf{s}$. These clauses are strongly related to point 1 in \Cref{def:common-expertise}. Again, abducible clauses are shared by all members of the team, since they all must be able to infer $\mathcal{S}^i$ from an arbitrary view $s^i$. For example:
\begin{lstlisting}[language=Prolog]
abducible(has_card_colour(P,S,C1)) :-
    player(P), slot(S), colour(C1), colour(C2), C2\==C1,
    not has_card_colour(P,S,C2), not ~has_card_colour(P, S, C1).
\end{lstlisting}
        indicates that a player may have a card of colour \texttt{C1} only if it is not known that they have a card of a different colour \texttt{C2} at that slot, nor is it explicitly stated that they do not have a card of colour \texttt{C1}.\footnote{We distinguish strong negation ($\sim$\texttt{Fact}) and negation as failure (\texttt{not Fact}). In epistemic logic notation, they are expressed as $K_i [\sim \phi]$ and $\sim K_i \phi$, respectively.}
        \item {\em Action selection clauses}: a set of clauses with annotated head \texttt{action(Agent, Action) [priority(N)]}, where \texttt{Agent} is an element in $G$, \texttt{Action} is an element in $\mathcal{A}$ and \texttt{N} is a number. Action selection clauses indicate what action to perform given the current observation that $i$ makes of the environment, and hence implement the team strategy function $\mathsf{Str}$. Action selection clauses have a particular feature: they are sorted according to the \texttt{priority(N)} annotation. When deciding on what action to take, agents consider action selection rules from lowest to highest priority, as detailed in \Cref{subsec:action-selection}. According to the common team strategy assumption in \Cref{def:common-expertise}, the action selection clauses are shared by all agents. For example:
\begin{lstlisting}
action(P, play_card(S)) [priority(N)] :-
    player_turn(P), has_card_color(P, S, C),
    has_card_rank(P, S, R), playable(C, R).
\end{lstlisting}
        indicates that the agent whose turn it is to move plays a safely playable card.
        \item {\em Abductive impossibility constraints} (AICs) are a set of clauses whose head has the annotated ground atom \texttt{imp [source(abduction)]}. This annotation serves to distinguish it from domain-related constraints, which have the same structure. AICs are used to integrate abductive explanations into the agent's program. Details about the generation and handling of AICs are provided in \Cref{subsec:explanations}. AICs are not shared across agents, as they are the result of an internal cognitive process. For example, in the example in \Cref{subsec:hanabi-game} (where Alice hints ``red'' to Cathy), Cathy would derive the following clause:
\begin{lstlisting}
imp [source(abduction)] :- ~has_card_rand(cathy,5,4).
\end{lstlisting}
    \end{itemize}
\end{itemize}

Now that we have presented the peculiarities of the domain and the components of the agents' programs, we explain next how agents make use of them when interacting. We split the exposition into three parts: (1) the abduction task that agents are faced with upon perceiving someone's action; (2) the refinement and assimilation of abductive explanations into their own knowledge base; and (3) the action selection process leveraging assimilated abductive explanations.

\subsection{The Abduction Task}\label{subsec:abduction-task}
At the current state $\mathbf{s}$, an \emph{acting agent} denoted by $j$, operating with logic program $T_j$, selects and performs action $a_j$. Denote an \emph{observer} agent by $i$, operating with logic program $T_i$. $i$, upon perceiving $j$ performing $a_j$, seeks to infer what knowledge $j$ was relying upon to select it. To do so, $i$ must switch to $j$'s perspective and not work with his own program $T_i$, but with the program that they approximate $j$ is working with at $\mathbf{s}$, which we denote by $T_{i,j}$ and define as:
\begin{equation}\label{eq:Tij}
\begin{split}
    T_{i,j} =& \{\phi \mid T_{i} \models \texttt{knows($j$,$\phi$)}\} \; \cup \\
    & \{h \; \texttt{:-} \; b_1, ..., b_n \in T_i \mid h \neq \texttt{imp [source(abduction)]}\}
\end{split}
\end{equation}
The first part in \cref{eq:Tij} states that the observer $i$ substitutes their view of the state $s^i$ by the view that they estimate $j$ has. This view, which we denote by $s^{i,j}$, is derivable from the ToM clauses. Hence, $s^{i,j}$ corresponds to the facts that $i$ knows that $j$ knows (in epistemic logic notation, $K_i K_j \phi$). The second part of \cref{eq:Tij} indicates that all the clauses in $i$'s program, with the exception of AICs, are carried over when $i$ adopts the perspective of $j$. This includes abducible clauses, which are necessary to infer the subset of states that $i$ thinks that $j$ believes to be possible, which we denote by $\mathcal{S}^{i,j}$.

Note that in this work we are assuming that ToM clauses are preserved when $i$ switches over to $j$'s perspective. However, this observation is not consequential to this work, because we only consider first-order Theory of Mind. The observer $i$ only invokes ToM clauses when switching from $T_i$ to $T_{i,j}$. Nonetheless, the switching of perspective can be extended to an arbitrary level of recursion:
\begin{equation}\label{eq:Tgeneral}
\begin{split}
	T_{i, j, ..., k, l} =& \{\phi \mid T_{i, j, ..., k} \models \texttt{knows}(l, \phi)\} \; \cup \\
	& \{h \texttt{:-} b_1, ..., b_n \in T_{i, j, ..., k} \mid h \neq \texttt{ic [source(abduction]}\}
\end{split}
\end{equation}
For example, $i$ could engage in second-order ToM by simulating the view that they know that $j$ knows that $k$ knows, i.e. $T_{i,j,k}$. In particular, it may be the case that $k=i$ ($T_{i,j,i}$), meaning that $i$ tries to see the world as $j$ thinks that $i$ is perceiving it. To generate $T_{i,j,k}$, then, agent $i$ would need to invoke ToM clauses from $T_{i,j}$, and hence possibly assume that $j$ is operating with the same ToM clauses as they are. Nevertheless, for the scope of this paper, it is not necessary to make such an assumption, as we do not go any further than first-order ToM.

Back to the main track of this work, the switch from program $T_i$ to $T_{i,j}$ corresponds to the observer agent $i$ engaging in first-order Theory of Mind and adopting the perspective of the acting agent $j$. In our view, this corresponds more closely to ST than to TT (see \Cref{subsec:tom}), as $i$ is simulating what the perception of the environment is from the point of view of $j$. $T_{i,j}$ is the approximation that agent $i$ builds of $T_j$. In general, $T_{i,j}$ is an incomplete version of $T_j$, as there are usually some parts of $s^j$ that are inaccessible to $i$ (i.e. the atoms in $s^j \backslash s^i$) and hence, according to \cref{eq:MAS-axioms}, not present in $T_{i,j}$.

$T_{i,j}$, then, is the logic program that $i$ has to work with in order to infer the knowledge that could have led $j$ to select action $a_j$. In other words, in order to generate abductive explanations for observation $Q=\texttt{action}(j, a_j)$, $i$ has to adopt logic program $T_{i,j}$. However, we are still missing the set of abducible atoms to build a complete ALP theory. The set of abducibles must include all ground literals that could complement the facts in $s^{i,j}$ to reconstruct $\mathbf{s}$. We denote them by $A_{i,j}$ and define them as:
\begin{equation}\label{eq:abducibles-ij}
    A_{i,j} = \{\alpha \mid T_{i,j} \models \texttt{abducible}(\alpha)\}
\end{equation}
Again, \cref{eq:abducibles-ij} can be generalised if the observer agent is engaging in higher-order ToM:
\begin{equation}\label{eq:abducibles-general}
    A_{i,j, ..., k, l} = \{\alpha \mid T_{i, j, ..., k, l} \models \texttt{abducible}(\alpha)\}
\end{equation}

In summary, upon getting notice of action $a_j$, the observer agent $i$ simulates being in the position of the acting agent $j$ and computes abductive explanations for observation $Q=\texttt{action}(j, a_j)$ with ALP theory $\langle T_{i,j}, A_{i,j} \rangle$. Abductive explanations, then, can be computed with the abductive extensions of SLD(NF). The output is a set of explanations $\{\Phi_1, ..., \Phi_m\}$, each corresponding to a subset of ground literals from $A_{i,j}$, $\Phi_l = \{\phi_{l1}, ..., \phi_{ln_l}\}$. Next, we present how such explanations are integrated back into the observer knowledge base, and how they are updated.

\subsection{Assimilation of Abductive Explanations}\label{subsec:explanations}
The abductive explanations obtained in the previous step, $\{\Phi_1, ..., \Phi_m\}$ with $\Phi_l = \{\phi_{l1}, ..., \phi_{ln_l}\}$, $\forall l \in [1,m]$, are useful if the observer $i$ can utilise them for their own decision-making.  Hence, the abductive explanations have to be integrated into $i$'s original program $T_i$. For this to happen, some post-processing is necessary. The post-processing of abductive explanations consists of two steps:
\begin{enumerate}
    \item For every abductive explanation $\Phi_l$, uninformative atoms are removed:
    \begin{equation*}
        \Phi_l' = \{\phi_{li} \mid \phi_{li} \in \Phi_l \; \text{and} \; T_i \not\models \phi_{li}\}
    \end{equation*}
    Only the informative facts of an explanation are kept, i.e. those that could not be derived from the original knowledge base. If after the removal of uninformative atoms $\Phi_l'$ is empty, the explanation is dropped altogether.
    
    \item For every (informative) abductive explanation, check that it is not impossible according to $i$'s current knowledge base:
    \begin{equation*}
        T_i \; \cup \; \Phi_l' \not\models \texttt{imp}
    \end{equation*}
    Explanations that are found to be impossible are removed. This is the point where our choice to adopt the consistency view of abductive explanations comes across. Here, the impossibility constraints against which consistency is checked include \emph{all} the clauses in $T_i$ with head \texttt{imp}: both domain-related and abductive constraints derived from previous abductive reasoning cycles.
\end{enumerate}
Note that step 1 is not strictly necessary, as (im)possible explanations will remain so even after uninformative facts have been removed. However, it helps to keep explanations less redundant and more compact.

Once the abductive explanations have been refined into $\{\Phi_1', ..., \Phi_{m'}'\}$, with $\Phi_l' = \{\phi_{l1}', ..., \phi_{ln_l'}'\}$, $\forall l \in [1,m']$, they are all integrated into a single logical formula in disjunctive normal form (DNF):
\begin{equation}\label{eq:dnf}
    \bigvee_{l=1}^{m'} \left( \bigwedge_{k=1}^{n_l'} \phi_{lk}' \right)
\end{equation}
Note that, for this initial proposal of our agent model, all (refined) abductive explanations are considered, i.e. it is not the case that one is selected as the most likely, nor are the various explanations weighed according to some numerical probability. Such extensions are left for future work.

In order to integrate the DNF in \cref{eq:dnf} into $T_i$ as a clause, one should consider that, as the DNF must hold, its negation must be false. Hence, the negation of the DNF can be used to construct an additional \emph{abductive impossibility constraint} clause to be appended to $T_i$:
\begin{lstlisting}[language=Prolog, mathescape=true]
imp [source(abduction)] :-
    (~$\phi_{11}'$ | ... | ~$\phi_{1n_1'}'$) & ... & (~$\phi_{m1}'$ | ... | ~$\phi_{mn_m'}'$).
\end{lstlisting}
The \texttt{[source(abduction)]} annotation indicates that this clause is not domain-related but derived from an internal cognitive process, and hence not shared by other agents. For the time being, the observer agent $i$ does not keep track of what AICs are derived from whose actions. However, a possible extension to this work could, for example, consider the level of trust among agents. The observer $i$ could be willing to integrate an abductive explanation into $T_i$ depending on whether the level of trust on the acting agent $j$ is above some given threshold.

In summary, when $i$ has concluded, through abductive reasoning, that $j$ knows about some fact about the state of the system, $i$ immediately incorporates it (as an annotated AIC clause). In epistemic logic notation, this is expressed by the axiom:
\begin{equation}
    K_i K_j \phi \rightarrow K_i \phi
\end{equation}
Note that this is the logical equivalent to \cref{eq:MAS-axioms}.

Abductive explanations and their corresponding AICs need to be updated as the observer agent gains access to information that was previously hidden. For example, in the Hanabi game, a player learns about the identity of a card they were holding the moment they play or discard it. More generally, denote an incoming piece of information by $\psi$, and the DNF derived from any previous abduction process by $\delta$, with the structure of \cref{eq:dnf}. If the addition of $\psi$ to the agent's program $T_i$ makes $\delta$ derivable ($T_i \cup \psi \models \delta$), the whole explanation is rendered uninformative. Therefore, the associated AIC clause has to be removed from $T_i$.

\subsection{Action Selection}\label{subsec:action-selection}
The whole purpose of the abductive reasoning task outlined in \Cref{subsec:abduction-task,subsec:explanations} is to provide the observer agent with new information that may be useful during their decision-making; that is, when the observer agent $i$ becomes the acting agent. In this section, we propose a default action selection procedure that takes into account abductive explanations. However, as stated in \Cref{sec:intro}, this is not the main focus of this work and this default proposal is susceptible of further investigation.

Recall that action selection for any member of the team happens according to the action selection rules, a set of clauses ordered by priority with head \texttt{action(Agent, Action)} and whose body states the conditions that must hold in order for \texttt{Action} to be selected by \texttt{Agent}. Consider the subset of action selection clauses that apply to $i$, i.e. \texttt{action($i$, Action)}. Because of imperfect information, it may be the case that the actor does not possess the necessary knowledge to prove the body of an action selection clause true given their current perception of the state $s^i$. Rather, the agent should evaluate whether the action is to be selected in the set of states $\mathcal{S}^i$ that are compatible with their current view $s^i$.

\begin{algorithm}[t]
\caption{Action selection query}
\label{alg:action-selection}
\KwData{Set of action selection clauses for $i$, \texttt{action}$(i, a)$ \texttt{[priority($n$)] :-} $\beta_a$}
\KwResult{A selected action $a_i$, or None}
\ForEach{action selection clause in ascending order of priority}{
    $\{\beta^{sk}\} \gets$ Skolemised forms of the rule body $\beta_a$\;
    \ForEach{Skolemised form $\beta_j^{sk} \in \{\beta^{sk}\}$}{
        $\{\beta^{tot}\} \gets$ instances of $\beta_j^{sk}$ that are compatible with the \texttt{imp} clauses\;
        \lIf{$T_i \cup \beta_i^{tot} \models \texttt{action}(a_i)$ for all $\beta_i^{tot} \in \{\beta^{tot}\}$}{
            \Return $a_i$
        }
    }
}
\Return None
\end{algorithm}

Depending on the domain at hand, the size of $\mathcal{S}^i$ may be prohibitively large to store, update, and loop over when evaluating action selection clauses. Instead of explicitly storing $\mathcal{S}^i$ and updating it during the abductive reasoning task, we need to consider only the features of a state that are relevant for selecting an action. The pseudocode for this computation appears in \Cref{alg:action-selection}.

\Cref{alg:action-selection} loops over the action selection clauses that apply to the now actor $i$ in ascending order of priority. For the clause under examination, all the Skolemised forms of the rule body are generated (line 2). Skolemised forms are computed as follows: whenever a subgoal of the rule body cannot be proven by the current knowledge base, its free variables are substituted by Skolem constants, i.e. constants that have not been previously encountered. This step is analogous to the computation of abductive explanations, where even if a subgoal cannot be proven, it may be added to the abductive explanation being constructed, as long as some instance of it is in the set of abducible atoms $A$. In general, for every action selection rule, several Skolemised forms of the rule body are derived.

Then, for every Skolemised form, its possible ``total'' instances are computed (line 4). This means that in the Skolemised form, Skolem constants have to be removed. Every time a literal is encountered that contains a Skolem constant, it is substituted by an {\em abducible} atom that matches with it (by treating Skolem constants as free variables). Abducible ground literals derived from $A_i$, see \cref{eq:abducibles-general}, are precisely the facts that may complement $s^i$ to build a complete representation of $\mathbf{s}$. However, note that we need only complement $s^i$ to the extent that it contains enough information to query the action selection clause under examination. Hence, \Cref{alg:action-selection} does not generate complete descriptions of the possible states in $\mathcal{S}^i$, but only the strictly necessary portions.

Not all potential instances of a Skolemised form are kept for further querying. Only those that are compatible with the impossibility constraints in $T_i$, i.e. $T_i \cup \beta_i^{tot} \not\models \texttt{imp}$. This includes both domain-related and, more importantly, abductive impossibility constraints. It is at this step that the abductive reasoning that the agent has previously engaged in pays off. The expectation is that the set of possible instances of a Skolemised form, $\{\beta^{tot}\}$, is smaller if AICs are considered, compared to the potential instances that would be compatible with domain-related constraints alone.

Finally, if every possible instance of the same Skolemised form leads to the same action being selected, that action is returned (line 5). This approach advocates for a totally safe action selection mechanism, as, for the time being, no quantification of uncertainty nor thresholds over such uncertainty are considered.

Although we only consider the action selection procedure of \Cref{alg:action-selection} in this paper, further work could extend the possibilities of an agent's action selection procedure based on their personality. For example, an agent may have a preference for action selection clauses that return an action given only their current knowledge, before considering additional inaccessible knowledge as in \Cref{alg:action-selection}.

\section{Results and Discussion}\label{sec:results}
We have implemented the agent model presented in \Cref{sec:combining} for the Hanabi domain in Jason \cite{Bordini2007a}, an agent-oriented programming language based on the BDI architecture.\footnote{ \url{https://github.com/nmontesg/hanabdi}} Jason allows for literal annotation (such as \texttt{imp [source(abduction)]} for AIC clauses) and custom Knowledge Query and Manipulation Language (KQML) performatives for agent communication, which we use to implement a specialised \texttt{publicAction} performative for agents to publicly announce their selected actions. The abduction task is performed by a meta-interpreter that does not fail unproven subgoals, but adds them to the abductive explanation under construction as long as they unify with an abducible atom.

As for the strategy being followed by the team, Hanabi has a vibrant community of online players who have gathered a set of ``conventions'' to follow during game play.\footnote{\url{https://hanabi.github.io/}} We have taken inspiration from these to design our action selection rules. Yet, for this preliminary work we stick to a fairly simple strategy (leaving out special moves such as ``prompts'' and ``finesses'').

\begin{figure}[t]
	\centering
	\begin{subfigure}{0.45\textwidth}
		\centering
		\includegraphics[width=\textwidth]{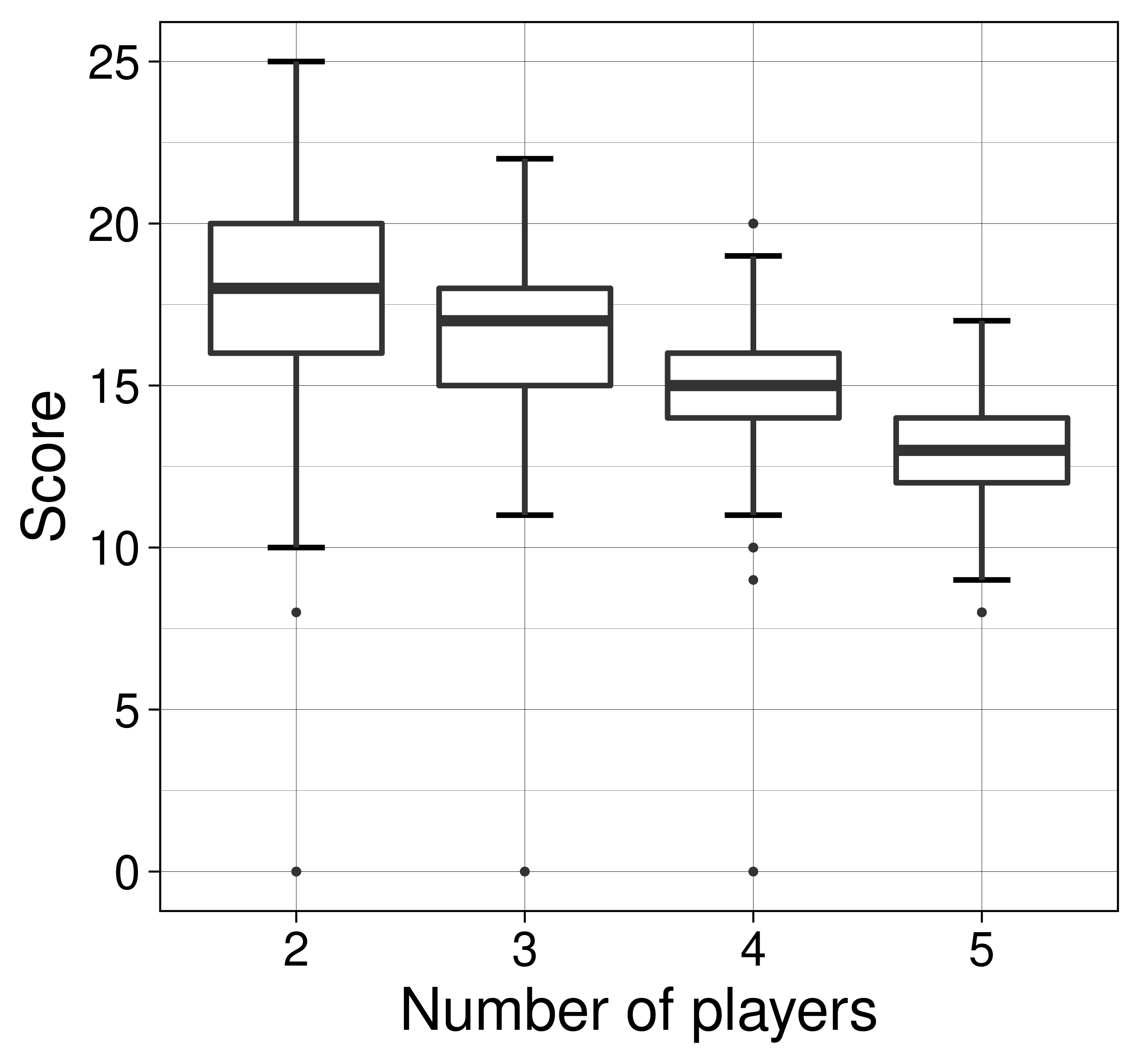}
		\label{subfig:score}
	\end{subfigure}
	\hfill
	\begin{subfigure}{0.45\textwidth}
		\centering
		\includegraphics[width=\textwidth]{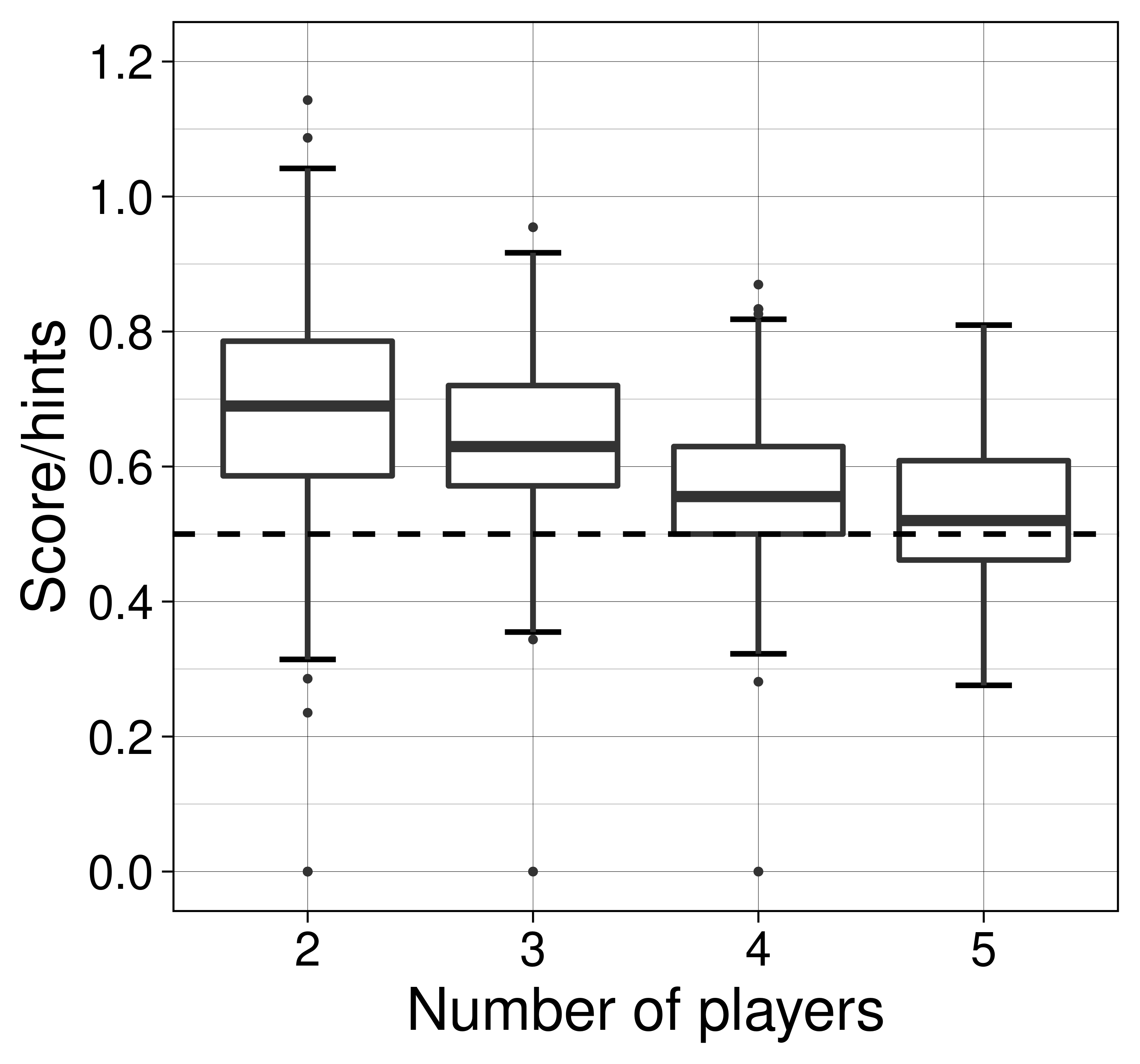}
		\label{subfig:efficiency}
	\end{subfigure}
	\caption{Results for the score (left) and communication efficiency (right) for our agent model applied to the Hanabi domain, for teams of 2 to 5 players. Every box contains data on 500 runs with different random seeds. The dashed line on the efficiency plot indicates the reference of 2 hints per score point.}
	\label{fig:results}
\end{figure}

We present a summary of our results in \Cref{fig:results}. We examine the performance of our team of ToM agents in terms of two metrics. The first is the obvious absolute team score (\Cref{subfig:score}), which is used by researchers involved in Hanabi AI as the standard indicator of performance. Our results do not match to the current state of the art (with average scores of up to 24.6 \cite{Lerer2020}). However, it should be noted that the work we present here is not concerned with the computation of an optimal playing policy, instead our concern is to provide a domain-independent agent model for cooperative domains. Predictably, the strategic conventions we have encoded can be fine-tuned and, potentially, optimised for the cognitive machinery our agents are endowed with.

The second performance metric we use is the {\em communication efficiency}, which we define as the ratio between the final score obtained and the number of hints provided throughout the game. This metric provides an indication of how effective is the team at converting exchanged information (hints) into utility (score points). To the best of the authors' knowledge, this metric has not been reported in any previous Hanabi AI work. Naively, a lower bound for communication efficiency is $\tfrac{1}{2}$, as two hints are required to completely learn about a card's identity (colour and rank) to safely play it. Nevertheless, this is a soft bound, due to correlations in colour and rank between different cards, or players being reiterated hints on cards they already have information on to prompt them to play. Still, our results in \Cref{subfig:efficiency} show that, for all team sizes, the communication efficiency falls above the $\tfrac{1}{2}$ mark for over half of the games. Just as the team strategy can be potentially optimised for the absolute score, it can also be automatically fine-tuned for the communication efficiency.

\section{Conclusions}\label{sec:conclusions}
In this work, we have presented a innovative agent model combining Theory of Mind and abductive reasoning for cooperation. In our framework, agents are able to understand their peers' motivations and hence enlarge their own imperfect information on the state of the environment. This model has been proven successful for the cooperative game of Hanabi, which offers an excellent test-bed to assess the performance of techniques for modelling others for teamwork.

Future work around this preliminary agent model should look into its generality (how far can the assumptions in \Cref{def:common-expertise} be relaxed while keeping the model sound); the optimisation of the team strategy given the proposed reasoning scheme; and its extension for \emph{ad hoc} teamwork, where agents autonomously tune their strategy to coordinate with previously unknown peers. This would require to relax the {\em common group strategy} assumption of the common expertise domains we have defined in this work, and adapt the agent model accordingly, possibly by incorporating other techniques from the goal and intention recognition literature.

\subsubsection*{Acknowledgements} This work has been supported by the EU WeNet project (H2020 FET Proactive project \#823783), the EU TAILOR project (H2020 \#952215), and the RHYMAS project (funded by the Spanish government, project \#PID2020-113594RB-100).


\bibliographystyle{splncs04}
\bibliography{references}

\end{document}